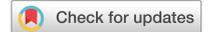

OPEN

# UBVRI night sky brightness at Kottamia Astronomical Observatory

Mohamed F. Aboushelib[1✉], A. B. Morcos[1], S. Nawar[1], O. M. Shalabiea[2,3] & Z. Awad[2]

Photoelectric observations of night sky brightness (NSB) at different zenith distances and azimuths, covering all the sky, at the Egyptian Kottamia Astronomical observatory (KAO) site of coordinates ϕ = 29° 55.9′ N and λ = 31° 49.5′ E, were done using a fully automated photoelectric photometer (FAPP). The Bessel wide range system (UBVRI) is used for the first time to observe NSB for three consecutive nights (1–3 August, 2022) under good seeing conditions after the moon sets. The deduced results were taken in photons and converted into mag/arcsec$^2$. The average zenith sky brightness for U, B, V, R and I filters are found to be 20.49, 20.38, 19.41, 18.60 and 17.94 mag/arcsec$^2$ respectively. The average color indices (U–B), (B–V), (V–R) and (R–I), at the zenith are detected to be 0.11, 0.98, 0.81 and 0.66, respectively. We plotted the isophotes of the sky brightness at KAO in U, B, V, R and I colors (filters) and determined both the average atmospheric extinction and sky transparency through these UBVRI filters. The atmospheric and other meteorological conditions were taken into our consideration during the observational nights. The results of the current study illustrate the main impact of the new cities built around KAO on the sky glow over it, and which astronomical observations are affected.

The night sky is not totally dark, but its brightness is the sum of different radiations resulting from different phenomena or sources. During moonless nights away from auroral regions, the total observed night sky brightness (NSB) is the sum of natural and artificial light sources. Natural sources include the effects of airglow, zodiacal light, integrated starlight, diffuse galactic light, extragalactic background light, solar activity and meteorological conditions while light pollution is the artificial source[1–5]. The brightness of the sky varies by changing time (diurnal, seasonal) and position. Measuring NSB indicates the artificial light contribution and shows the change in the brightness and color of the sky that represent one of the main parameters in choosing suitable sites for establishing astronomical observatories[6].

Measuring NSB began since 1899 by Burns[7] followed by Newcomb[8]. They used a visual observational method, by comparing the brightness of the sky to an out of focus disk of a star, through a telescope. Other studies of NSB involved using photographic plates[9,10], and visual photometers[11–16]. Rudnick[17] used a camera supported by blue and red filters to measure NSB at different zenithal distances. Also photoelectric photometers were used with different filters in several studies[5,18–28]. Since the beginning of the 2000s, CCD cameras, equipped with lenses or telescopes[29–31], became widely used to observe NSB. Patat[32] followed by Krisciunas et al.[33] used CCD cameras supported by UBVRI filters to measure zenith NSB.

In the present work and for the first time, we measure the NSB and the color distribution at KAO using a customized fully automated photoelectric photometer (FAPP), supported by Bessel UBVRI filters to cover the whole sky for different azimuths and zenith distances.

## KAO site and instrument

Observations were carried out at KAO in Egypt which is located at 29° 55′ 48″ N & 31° 49′ 30″ E about 80 km East from Cairo city as shown in Fig. 1. KAO lies at an elevation of about 480 m above the sea level. The observations were carried out on three consecutive nights, August 1st, 2nd and 3rd, 2022. The KAO has been operating since 1964 with two telescopes, the large telescope which has a mirror of 74 inches diameter and the small telescope of a 14 inches diameter mirror. Since 1990s, new cities have been constructed near the observatory site and from the beginning of 2016; the New Administrative Capital city has been constructed 13 km away west of the observatory.

The study of the influence of constructing these new cities on the efficiency of astronomical observations at KAO and the effect of light pollution from them on NSB, are the main concern of this work. Recent evaluation for the variation of NSB over KAO is urgently needed because the last study conducted was in 1995 and used a semi-automated photoelectric photometer with three (BVR) filters[25,34,35].

[1]National Research Institute of Astronomy and Geophysics, Helwan, Cairo, Egypt. [2]Astronomy, Space Science and Meteorology Department, Faculty of Science, Cairo University, Giza, Egypt. [3]Faculty of Navigation Science and Space Technology, Beni-Suef University, Beni-Suef, Egypt. ✉email: m.f.aboushelib@nriag.sci.eg

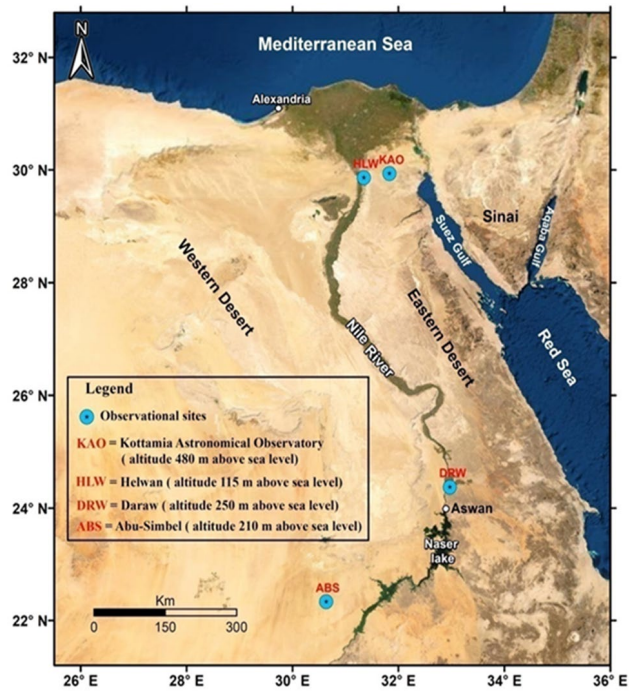

**Figure 1.** A map of Egypt showing different observational sites including the Kottamia Astronomical Observatory (KAO).

We used a customized fully automated photoelectric photometer (FAPP) with multiple filters. The device had been tested and calibrated by a team of professional night sky researchers at the Astronomy Department at the National Research Institute of Astronomy and Geophysics (NRIAG) and the manufacturing company. The manual of the device is under publication, since it is not on-shelf device.

FAPP is controlled by a customized software that controls the mount, the hole opening, the filter wheels, the scan mode, scan coordinates, the range of Alt-Az and the step of the scan. The scan step can be chosen to be either in the direction of both the Alt-Az or separate in each direction. The device is also supported by a software for storing data and scripts to calculate the NSB and Twilight.

The main photometer box Fig. 2, contains a photomultiplier tube (PMT) of the type 9798B series manufactured by ET Enterprises Limited, The UK[36]. The PMT is a photocathode that is sensitive to radiation in ultraviolet to infra-red range at high gain and high stability even when its operating voltage is low.

The minimum and maximum dark currents are $2 \times 10^{-9}$ and $20 \times 10^{-9}$ amperes for the lowest and highest voltages respectively. Inside the photometer box, there is a wheel that contains five wide band Bessel filters (UBVRI); their range and effective wavelengths are shown in Fig. 3.

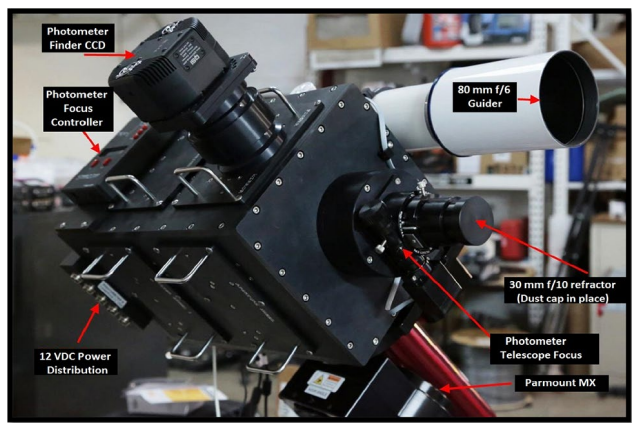

**Figure 2.** The fully automated photoelectric photometer (FAPP) main box with labels to indicate the main components of the instrument.

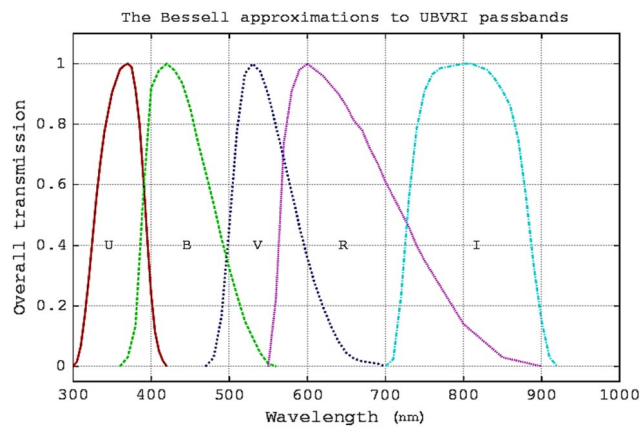

**Figure 3.** Transmission curves of the five wide band Bessel filters.

There is another wheel which contains 8 different holes with different diameters (plus one more hole of the same diameter as the 8th one). These holes are used to control the intensity of incident light that reaches the photomultiplier tube. The diameters of these holes from 1 to 8 are 1, 2, 3, 4, 6, 8, 9 and 12 mm, respectively, with an extra 12 mm hole.

The area of the sky corresponding to each hole is computed theoretically by using the well-known relation:

$$Area\ of\ the\ sky = \left(area\ of\ the\ \frac{hole}{f^2}\right) \times (10^4/3.046)\ \text{square degree} \quad (1)$$

where the focal length ($f$) of the objective lens equals 300 mm

The signal from phototube is sent to AD8 amplifier. The AD8 amplifier-discriminator is a compact low noise high gain electronics module designed by ET Enterprises Limited, The UK. It is used in PMT photon counting system to generate Transistor–Transistor Logic (TTL) output pulses. The TTL output pulses reach the multi-channel Scaler/Counter Timer module (MCS-CT3) pulse counting instrument which is a compact electronics module to record and store pulses as a function of time.

## Measurements and results

FAPP is adjusted to scan all the sky with 10° steps in both Alt and Az directions during the observation period after the end of the astronomical twilight and the setting of the moon using UBVRI filters. The time of Moonrise and set, the direction, the Moon phase and illumination at each observing night are listed in Table 1. These data are taken into consideration to eliminate all the effects of moonlight on our observations.

Throughout the interval of observations, we fixed the device observing hole to be hole 8 which has a diameter of 12 mm. This opening corresponds to a field of view of 4.1234 sq. degree, calculated by Eq. (1). The exposure time was fixed to 3s. The working voltage of the phototube is kept constant at 700 V.

It is worth mentioning that four full scans for each filter were done per night. The photometer measures the photon flux in units of photon/cm²/second/steradian (photon/cm²/s/sr). This unit must be converted into magnitude/arcsecond² (mag/arcsec²).

The sky brightness can be changed from $I(S_{10})$ into $I(mag/arcsec^2)$ for a certain wavelength by using the following relation[22]

$$I(\text{mag/arcsec}^2) = \left(-2.5 \times Log I(S_{10})\right) + 27.78 \quad (2)$$

where $S_{10}$ is the brightness of a 10's magnitude star which is defined by the following equation

$$S_{10} = 2.512^{(10-C)} \text{ and } C = Log I_0/0.4 \quad (3)$$

| Date | Moonrise | | Moonset | | |
|---|---|---|---|---|---|
| | Time UT | Az degree (°) | Time UT | Az degree (°) | Illumination (%) |
| 1-Aug | 6:21 | 80 | 19:15 | 276 | 13.40 |
| 2-Aug | 7:17 | 87 | 19:43 | 270 | 21.30 |
| 3-Aug | 8:15 | 94 | 20:13 | 263 | 30.60 |

**Table 1.** Moon (rise and set, direction and phase) each night.

where $LogI_0$ can be found by plotting ($LogI$) against the air mass ($X$). The plot is a straight line and its slope gives the optical thickness and its intersection with log I axis at $X = 0$ gives $LogI_0$. Or simply by solving a family of equations in the form

$$LogI = LogI_0 - KX \qquad (4)$$

where the airmass ($X$) is given by the following expression

$$X = \sec z - 0.0018167 \times (\sec z - 1) - 0.002875 \times (\sec z - 1)^2 \qquad (5)$$

where $z$ is the zenith distance

A simple script was added to the FAPP software to do these mathematical computations. Four standard stars (Alpha Aquilae, Vega, Alpha Cygni and Alpha Scorpii), were observed through each filter, during each night at different altitudes to determine the atmospheric extinction. All the readings from the five filters were converted into $S_{10}$ and mag/arcsec$^2$. The average readings ($U_{avg}$, $B_{avg}$, $V_{avg}$, $R_{avg}$ and $I_{avg}$) for the three observing nights and the color indices for different altitudes from 0° to 90° are included (in Tables S-1 through S-10) in the supplementary file. The following table (Table 2) shows the average NSB in different colors, and the color indices for the three nights at Alt 90°.

The isophotes of NSB for all filters are drawn against Alt and Az in degrees in the following figures (Figs. 4, 5, 6, 7 and 8).

| Az degrees | Alt degrees | $U_{avg}$ mag/arcsec2 | $B_{avg}$ mag/arcsec2 | $V_{avg}$ mag/arcsec2 | $R_{avg}$ mag/arcsec2 | $I_{avg}$ mag/arcsec2 | U–B | B–V | V–R | R–I |
|---|---|---|---|---|---|---|---|---|---|---|
| 0 | 90 | 20.44 | 20.26 | 19.27 | 18.57 | 17.84 | 0.18 | 0.99 | 0.70 | 0.73 |
| 10 | 90 | 20.29 | 20.44 | 19.38 | 18.68 | 17.93 | −0.15 | 1.06 | 0.70 | 0.75 |
| 20 | 90 | 20.24 | 20.47 | 19.37 | 18.67 | 18.02 | −0.23 | 1.10 | 0.70 | 0.65 |
| 30 | 90 | 20.46 | 20.47 | 19.48 | 18.70 | 17.68 | −0.01 | 0.99 | 0.78 | 1.01 |
| 40 | 90 | 20.44 | 20.51 | 19.47 | 18.78 | 17.98 | −0.06 | 1.04 | 0.69 | 0.79 |
| 50 | 90 | 20.53 | 20.03 | 19.47 | 18.77 | 18.11 | 0.49 | 0.57 | 0.70 | 0.66 |
| 60 | 90 | 20.55 | 20.08 | 19.41 | 18.70 | 17.99 | 0.46 | 0.68 | 0.71 | 0.71 |
| 70 | 90 | 20.26 | 20.28 | 18.36 | 17.55 | 17.89 | −0.02 | 1.92 | 0.81 | −0.34 |
| 80 | 90 | 20.45 | 20.40 | 19.38 | 18.27 | 17.30 | 0.05 | 1.02 | 1.12 | 0.97 |
| 90 | 90 | 20.52 | 20.50 | 19.04 | 18.29 | 17.45 | 0.01 | 1.46 | 0.75 | 0.84 |
| 100 | 90 | 20.58 | 20.54 | 19.53 | 18.82 | 18.15 | 0.04 | 1.01 | 0.71 | 0.67 |
| 110 | 90 | 20.59 | 20.54 | 19.53 | 18.84 | 18.17 | 0.05 | 1.01 | 0.70 | 0.67 |
| 120 | 90 | 20.60 | 20.54 | 19.54 | 18.82 | 18.18 | 0.06 | 1.00 | 0.72 | 0.64 |
| 130 | 90 | 20.59 | 20.54 | 19.53 | 18.80 | 18.18 | 0.06 | 1.00 | 0.73 | 0.62 |
| 140 | 90 | 20.61 | 20.54 | 19.53 | 18.80 | 18.16 | 0.07 | 1.01 | 0.73 | 0.63 |
| 150 | 90 | 20.59 | 20.54 | 19.53 | 18.80 | 18.15 | 0.05 | 1.01 | 0.74 | 0.64 |
| 160 | 90 | 20.61 | 20.53 | 19.54 | 18.80 | 18.16 | 0.07 | 1.00 | 0.74 | 0.64 |
| 170 | 90 | 20.59 | 20.54 | 19.55 | 18.80 | 18.17 | 0.04 | 1.00 | 0.75 | 0.63 |
| 180 | 90 | 20.49 | 20.43 | 19.43 | 18.68 | 17.97 | 0.06 | 1.00 | 0.75 | 0.71 |
| 190 | 90 | 20.56 | 20.47 | 19.45 | 18.70 | 17.97 | 0.09 | 1.02 | 0.75 | 0.72 |
| 200 | 90 | 20.58 | 20.48 | 19.47 | 18.71 | 18.03 | 0.09 | 1.01 | 0.76 | 0.68 |
| 210 | 90 | 20.57 | 20.52 | 19.52 | 18.75 | 18.09 | 0.05 | 1.00 | 0.77 | 0.65 |
| 220 | 90 | 20.53 | 20.54 | 19.48 | 18.74 | 18.06 | −0.01 | 1.06 | 0.74 | 0.67 |
| 230 | 90 | 20.59 | 19.24 | 19.54 | 18.79 | 18.14 | 1.35 | −0.30 | 0.75 | 0.64 |
| 240 | 90 | 20.59 | 19.03 | 19.54 | 18.78 | 18.17 | 1.56 | −0.51 | 0.76 | 0.62 |
| 250 | 90 | 20.00 | 20.42 | 19.25 | 17.46 | 17.87 | −0.41 | 1.16 | 1.79 | −0.41 |
| 260 | 90 | 20.05 | 20.37 | 19.18 | 17.40 | 16.41 | −0.32 | 1.19 | 1.78 | 0.99 |
| 270 | 90 | 20.44 | 20.39 | 18.92 | 17.58 | 16.54 | 0.05 | 1.47 | 1.34 | 1.05 |
| 280 | 90 | 20.51 | 20.52 | 19.49 | 18.79 | 18.13 | −0.02 | 1.03 | 0.70 | 0.65 |
| 290 | 90 | 20.49 | 20.52 | 19.49 | 18.79 | 18.13 | −0.03 | 1.03 | 0.70 | 0.66 |
| 300 | 90 | 20.50 | 20.52 | 19.49 | 18.79 | 18.14 | −0.01 | 1.03 | 0.70 | 0.65 |
| 310 | 90 | 20.53 | 20.51 | 19.49 | 18.80 | 18.14 | 0.02 | 1.02 | 0.70 | 0.66 |
| 320 | 90 | 20.54 | 20.51 | 19.50 | 18.80 | 18.15 | 0.03 | 1.01 | 0.70 | 0.65 |
| 330 | 90 | 20.57 | 20.51 | 19.52 | 18.82 | 18.15 | 0.07 | 0.99 | 0.70 | 0.66 |
| 340 | 90 | 20.53 | 20.51 | 19.51 | 18.80 | 18.14 | 0.02 | 1.01 | 0.70 | 0.66 |
| 350 | 90 | 20.59 | 20.52 | 19.52 | 18.82 | 18.16 | 0.07 | 1.00 | 0.70 | 0.66 |
| 360 | 90 | 20.52 | 20.29 | 19.30 | 18.60 | 17.88 | 0.22 | 1.00 | 0.70 | 0.71 |

**Table 2.** The average NSB in different colors and the color indices for the three nights at Alt 90°

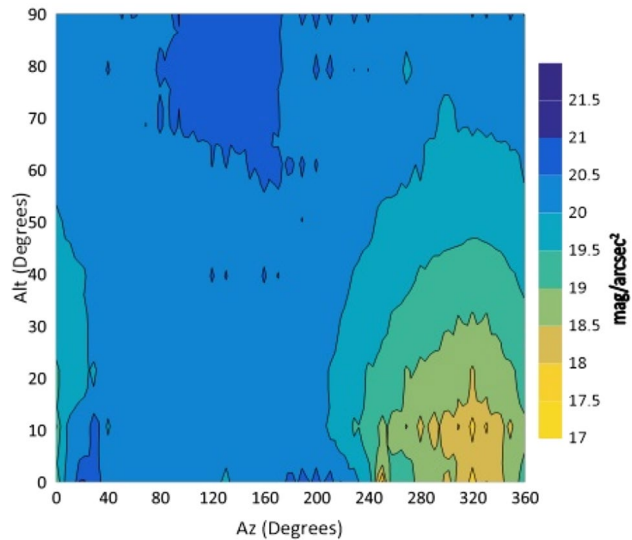

**Figure 4.** The night sky brightness (NSB) in the U filter at different Altitudes and Azimuths.

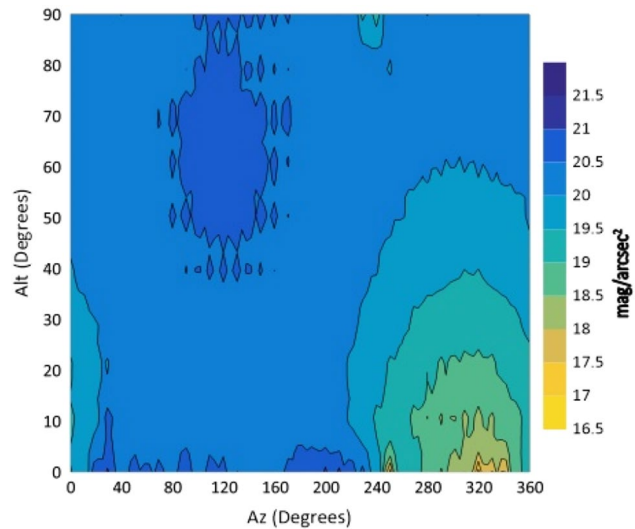

**Figure 5.** The night sky brightness (NSB) in the B filter at different Altitudes and Azimuths.

From the isophotes, in all filters, it is clear that the maximum brightness occurs at Azimuth angles in the range 240°–260° and altitudes below 10°. When the Azimuth ranges from 280° to 360°, the sky glows and it is graduated from 14.4 to 15.9 mag/arcsec$^2$ at Alt between 0° and 30°. At the Az range from 0° to 240°, NSB graduated from 18 to 21 mag/arcsec$^2$ for BVR filters. The sky glow extends from the horizon to the zenith with different degrees. The minimum brightness is located at Az range between 80° and 170° to the zenith (Alt 90°). The average zenith brightness, in the UBVRI filters, is 20.49, 20.38, 19.41, 18.60 and 17.94 mag/arcsec$^2$, respectively. The average color indices at the zenith are U–B = 0.11, B–V = 0.98, V–R = 0.81 and R–I = 0.66. The atmospheric extinction and sky transparency of the UBVRI filters were determined for the period of observations and the results are listed in Table 3.

The transparency was calculated from Beer–Lambert law[37] as

$$T = e^{-\tau} \qquad (6)$$

where τ = 2.303 K

Table 4 represents the aerosol optical depth (AOD), precipitable water vapor (PWV) and cloud fraction (CF) for the three observational days from satellite data (Terra/MODIS) with level 3 MODIS with 1 degree spatial resolution (29° N–30° N and 31° E–32° E) and daily temporal resolution. For the AOD, we chose the (Dark Target Deep Blue Combined Mean) which combines land and ocean of AOD at 550 nm. For the PWV (IR retrieval), we chose the Total Column: Mean of Level-3 QA Weighted Mean. Hence the PWV is the total amount of water

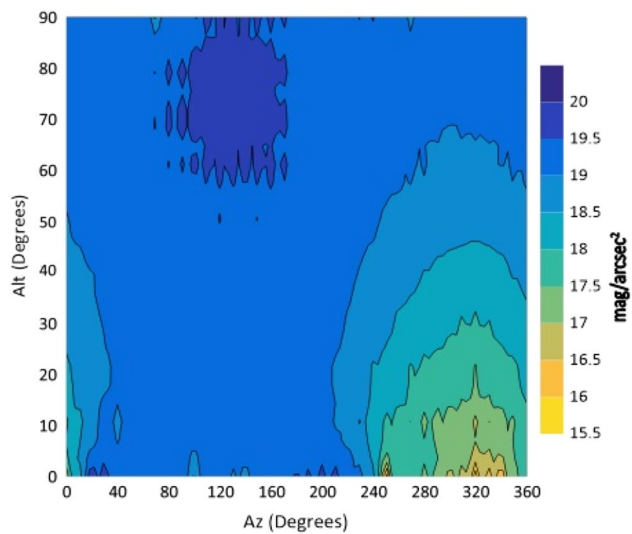

**Figure 6.** The night sky brightness (NSB) in the V filter at different Altitudes and Azimuths.

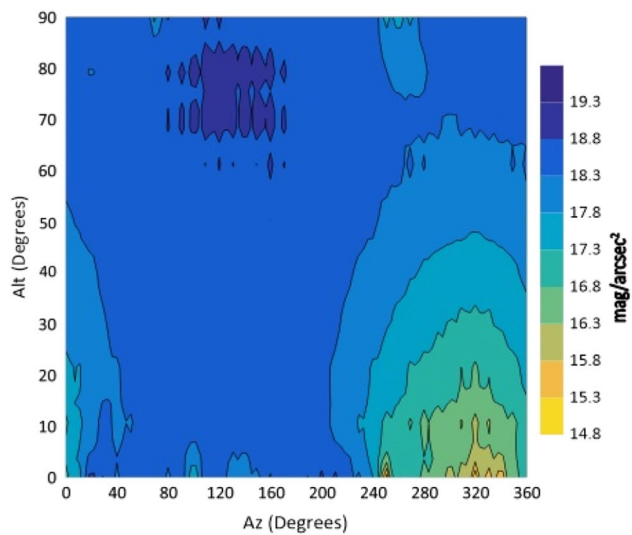

**Figure 7.** The night sky brightness (NSB) in the R filter at different Altitudes and Azimuths.

vapor in the atmosphere. For the CF, we chose the Cloud Fraction from Cloud Mask (count of lowest 2 clear sky confidence levels, cloudy and probably cloudy / total count): Mean of Daily Mean (MOD08_D3 v6.1). (Data was retrieved from https://giovanni.gsfc.nasa.gov/)[38]

The AOD data shows a presence of aerosol during the nights of observations; the highest was on August 1st and the lowest on August 2nd. The PWV increased gradually during the three nights to reach its maximum on August 3rd while cloud fraction was almost zero through the three nights. Aerosol and water vapor may increase the scattering of radiation which in turn leads to higher sky brightness values. The presence of the aerosol has a significant impact on the NSB, especially at rural and urban sites creating an artificial sky glow. Many studies have found a correlation between the atmospheric aerosol and the sky brightness[4,39–42]. The sky can get brighter at air polluted sites where the brightness depends on the amount of the aerosol and the distance to the light source. For near light sources, the impact is larger in which scattering increases and the sky brightness can vary by 10s%[39].

### Analysis of the errors on the results

The computed maximum (Max), minimum (Min) and mean sky brightness values in the UBVRI filters, together with both the standard error and the standard deviation are shown in Table 5, at altitudes of 0°, 30°, 60° and 90°. A visualization of these errors at Alt = 30° as a function of Az is illustrated in Fig. 9.

Variations in the sky brightness have been recognized for all filters. For instance, at an Alt of 30°, the sky brightness is the lowest for Az between 80° and 170° while it records the highest values at Az between 300° and 330°, for all filters. The difference between maximum and minimum sky brightness is the largest (4.10, 4.44, 4.40, 4.33 and 4.22 mag/arcsec$^2$) at Alt 0° and the lowest at Alt 60° (0.78, 0.53, 0.67, 0.66 and 0.61 mag/arcsec$^2$) for UBVRI filters, respectively.

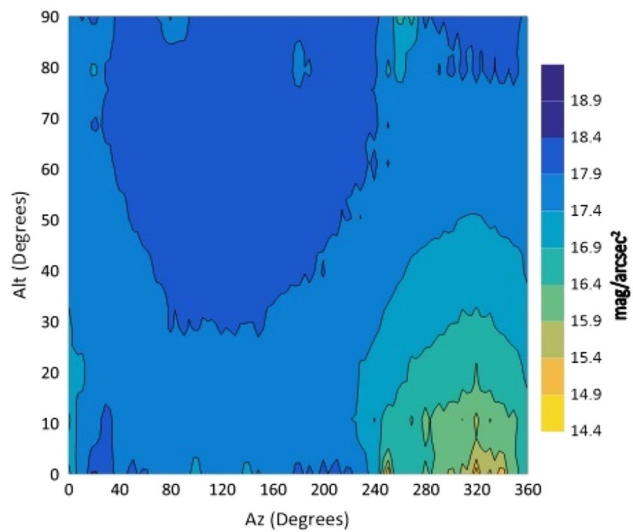

**Figure 8.** The night sky brightness (NSB) in the I filter at different Altitudes and Azimuths.

| Filter | Extinction | Transparency |
|---|---|---|
| U | 0.71 | 0.49 |
| B | 0.47 | 0.62 |
| V | 0.31 | 0.73 |
| R | 0.28 | 0.76 |
| I | 0.24 | 0.79 |

**Table 3.** Average atmospheric extinction and transparency of UBVRI during observation period of August 1st–3rd, 2022.

| Date | AOD | PWV (cm) | CF % |
|---|---|---|---|
| 1-Aug | 0.38 | 2.92 | 0 |
| 2-Aug | 0.24 | 3.14 | 0 |
| 3-Aug | 0.28 | 3.46 | 0 |

**Table 4.** The aerosol optical depth (AOD), the precipitable water vapor (PWV) in cm and the percentage of cloud fraction (CF) during observations.

The standard error within the results shows a decrease with the zenith distance reaching Alt 60° followed by a little increase for B, R and I filter at Alt 90°. The error is the highest at Alt 0° (zenith distance 90°) and the lowest at Alt 60° (zenith distance 30°), for most of the filters. At the horizon, the atmosphere can be hazy and dusty due to the presence of aerosol and water vapor which can alter the sky brightness. At Alt 30° and 60°, the error percentage was almost the same for all filters while at Alt 90°, the error was the lowest for the U filter and the highest for both the R and I filters. The resultant error is due to the fact that the sky brightness is not homogeneous in all directions. The inhomogeneity of the sky brightness results from the light pollution which is most significant towards the direction of the light source in addition to the atmospheric aerosol and water vapor. We may conclude that the standard error for the measurements of this study is small and can be acceptable.

A comparison between the observed NSB in this work (for the 5 filters) and the NSB observed at different observatory locations are listed in Table 6. The measured sky values in the current study are brighter than other compared sites. The light pollution from nearby cities has the most significant impact on the sky brightness. The elevation and atmospheric conditions have also an impact on the NSB. Compared to previous observations at KAO in 1995[35], the NSB increased, in the present work, by about 9.4%, 9.8% and 9.6% in the BVR filters.

| Filter | U | B | V | R | I |
|---|---|---|---|---|---|
| Alt 0° | | | | | |
| Max. brightness (mag/arcsec$^2$) | 17.22 | 16.79 | 15.58 | 14.80 | 14.46 |
| Min. brightness (mag/arcsec$^2$) | 21.32 | 21.23 | 19.97 | 19.13 | 18.68 |
| Mean (mag/arcsec$^2$) | 19.74 | 19.78 | 18.41 | 17.60 | 17.10 |
| Standard error | 0.17 | 0.22 | 0.22 | 0.21 | 0.19 |
| Standard deviation | 1.05 | 1.33 | 1.33 | 1.28 | 1.13 |
| Alt 30° | | | | | |
| Max. brightness (mag/arcsec$^2$) | 18.81 | 19.10 | 17.86 | 17.06 | 16.78 |
| Min. brightness (mag/arcsec$^2$) | 20.43 | 20.50 | 19.42 | 18.70 | 17.97 |
| Mean (mag/arcsec$^2$) | 19.81 | 19.94 | 18.83 | 18.08 | 17.52 |
| Standard error | 0.09 | 0.08 | 0.09 | 0.09 | 0.07 |
| Standard deviation | 0.57 | 0.49 | 0.55 | 0.57 | 0.41 |
| Alt 60° | | | | | |
| Max. brightness (mag/arcsec$^2$) | 19.87 | 20.06 | 18.93 | 18.21 | 17.60 |
| Min. brightness (mag/arcsec$^2$) | 20.65 | 20.59 | 19.61 | 18.87 | 18.20 |
| Mean (mag/arcsec$^2$) | 20.31 | 20.35 | 19.33 | 18.60 | 17.99 |
| Standard error | 0.04 | 0.03 | 0.04 | 0.04 | 0.03 |
| Standard deviation | 0.23 | 0.18 | 0.22 | 0.22 | 0.17 |
| Alt 90° | | | | | |
| Max. brightness (mag/arcsec$^2$) | 20.00 | 19.03 | 18.36 | 17.40 | 16.41 |
| Min. brightness (mag/arcsec$^2$) | 20.61 | 20.54 | 19.55 | 18.84 | 18.18 |
| Mean (mag/arcsec$^2$) | 20.49 | 20.38 | 19.41 | 18.60 | 17.94 |
| Standard error | 0.02 | 0.05 | 0.04 | 0.07 | 0.07 |
| Standard deviation | 0.14 | 0.33 | 0.23 | 0.41 | 0.41 |

**Table 5.** The max, min and mean measured NSB, the standard error and standard deviation of the UBVRI filters at different Alt.

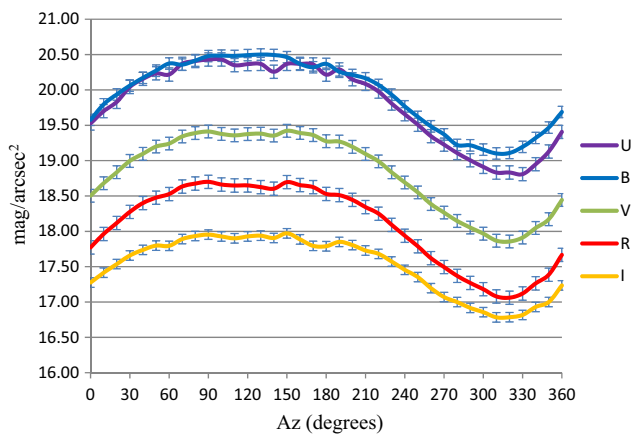

**Figure 9.** The variation of the average sky brightness in the UBVRI with different azimuth degrees at Alt 30°

## Conclusion

The FAPP is an accurate and speedy device with high facilities. It can be used not only in NSB measurements, but also in measuring the other component of sky, such as, the twilight, zodiacal light. We calculated the atmospheric extinction in each filter to convert brightness from photon/cm$^2$/s/sr into the standard unit of mag/arcsec$^2$. The calculated extinction coefficients are 0.71, 0.47, 0.31, 0.28 and 0.24; whereas the average zenith brightness is 20.49, 20.38, 19.41, 18.60 and 17.94 mag/arcsec$^2$ for the five UBVRI filters, respectively. The color indices are obtained to be U−B = 0.11, B−V = 0.98, V−R = 0.81 and R−I = 0.66.

The observations showed areas of high brightness, mostly at the west (at Az ranges 240°–260° and 280°–360°, but for Alt. less than 30°). The lowest sky brightness was found at Az between 80° and 170° at Alt between 60° and 90°. New cities increased the sky brightness over the KAO. The largest impact is due to the New Administrative Capital city which contaminates the observing site of the KAO at Az range between 240° and 360°. Badr city also has a significant impact within the Az range between 0° and 20° but less than that of the New Administrative city.

| Site | Elevation (meter) | NSB (mag/arcsec²) | | | | |
|---|---|---|---|---|---|---|
| | | U | B | V | R | I |
| Catania (Italy)[18] | 1735 | 21.37 | 21.13 | 20.8 | – | – |
| Mauna Kea (USA)[21] | 2800 | – | 22.87 | 21.91 | – | – |
| Calar Alto (Spain)[22] | 2168 | 22.39 | 22.86 | 22.01 | 21.36 | 19.25 |
| Paranal (Chile)[32] | 2635 | 22.3 | 22.6 | 21.6 | 20.9 | 19.7 |
| Kottamia (Egypt)[35] | 480 | – | 22.5 | 21.51 | 20.57 | – |
| Abu Simble (Egypt)[43] | 270 | – | 22.58 | 21.66 | – | – |
| Kottamia (Egypt) Current work | 480 | 20.49 | 20.38 | 19.41 | 18.6 | 17.94 |

**Table 6.** A comparison between the measured NSB at KAO (this work, 1st raw) and previous work with that at other observatories and observing sites. The elevation of each site is included.

In the U filter, the minimum brightness was found between Az 70°–160° at Alt > 50° with an average brightness of 20.58 mag/arcsec². In both the B and V filters, the minimum sky brightness was found between Az 80°–170° at Alt > 50° and 60° where the average brightness is 20.54 and 19.56 mag/arcsec², respectively. For the R filter the minimum brightness was found between Az 80°–170° at Alt ≥ 70° with average brightness of 18.84 mag/arcsec². In the I filter, the minimum brightness was found between Az 40°–220° at Alt ≥ 30° and most of the sky at Alt > 80. The presence of aerosol and water vapor increases the atmospheric extinction and hence affects the sky brightness. The aerosol increases the light scattering while water vapor increases the absorption of light. A significant dispersion was noticed for V, R and I filters during the three nights of observations. For all filters, the sky on August 1st was brighter than that on August 3rd. A direct correlation between the sky brightness and the AOD and an inverse relation with the PWV were found.

The values of the sky brightness are indicators of the sky brightness at different wavelengths and the color index is an indication of the color of the sky and hence the nature of the light sources that create the sky glow at the site. The color of the sky appears to shift towards the red which reflects the nature of most of light sources affecting the sky brightness at the observatory. The tendency of the color of the sky to be redder reflects the nature of a warm color light source. We know that most of the lamps installed in the New Administrative Capital are a mix between warm white light-emitting diode (LED) and high-pressure sodium (HPS) since at some sites under construction, HPS lamps are still used.

Observing the whole sky showed that large parts are affected by light pollution from nearby cities but at lower altitudes. However, there are some sky areas suitable for some types of observations. It is expected that with time, urban sprawl in the direction of the observatory, the increase in the intensity of luminescence and light pollution may lead to the inability to take photometric observations and lead to make the observatory not working efficiently.

## Data availability
The datasets used and/or analysed during the current study available from the corresponding author on reasonable request.

### Author contributions
M.F.A. carried out the observations and wrote the Abstract, Introduction, Method and results, Analysis of the errors on the results and Conclusion sections. A.B.M. wrote the KAO site and Instrument section. A.B.M., S.N., O.M.S. and Z.A. are supervisors and reviewed the manuscript.

### Funding
Open access funding provided by The Science, Technology & Innovation Funding Authority (STDF) in cooperation with The Egyptian Knowledge Bank (EKB).

### Competing interests
The authors declare no competing interests.

### Additional information
**Supplementary Information** The online version contains supplementary material available at https://doi.org/10.1038/s41598-023-43844-x.

**Correspondence** and requests for materials should be addressed to M.F.A.

**Reprints and permissions information** is available at www.nature.com/reprints.

**Publisher's note** Springer Nature remains neutral with regard to jurisdictional claims in published maps and institutional affiliations.